\documentclass[a4paper,UKenglish]{lipics-v2018}

\usepackage{graphicx}
\usepackage{amsfonts,amsmath,latexsym,amssymb}
\usepackage[utf8]{inputenc}
\usepackage[ruled]{algorithm}
\usepackage{algorithmicx}
\usepackage{algpseudocode}
\usepackage{scrextend}

\usepackage[T1]{fontenc} 
\usepackage{lmodern} 
\rmfamily 
\DeclareFontShape{T1}{lmr}{b}{sc}{<->ssub*cmr/bx/sc}{}
\DeclareFontShape{T1}{lmr}{bx}{sc}{<->ssub*cmr/bx/sc}{}

\definecolor{LinksColor}{rgb}{0,0.2,0.8}

\nolinenumbers

\graphicspath{{./FIGURES/}} 

\title{Information Gathering in Ad-Hoc Radio Networks}

\titlerunning{Information Gathering in Ad-Hoc Radio Networks}

\author{Marek Chrobak}{Department of Computer Science\\ University of California at Riverside}{}{}{}

\author{Kevin P. Costello}{Department of Mathematics\\ University of California at Riverside}{}{}{}

\author{Leszek G\k{a}sieniec}{Department of Computer Science\\ University of Liverpool}{}{}{}

\Copyright{The copyright is retained by the authors}

\authorrunning{M.~Chrobak, K.~Costello, L.~G\k{a}sieniec}

\subjclass{%
}


\funding{M.~Chrobak's research supported by NSF grant CCF-1536026.}

\newcommand{\lastcorrections}%
{{
 \begin{sloppypar}
    \baselineskip -0.2in
    \tiny\bf\noindent
last corrections:\\
\end{sloppypar}
}}

\newcommand{\margincomment}[1]%
    {{%
      \marginpar{{\tiny\begin{minipage}{0.5in}
                       \begin{flushleft}
                          {#1}
                       \end{flushleft}
                       \end{minipage}
                }}
    }}

\newcommand{\ignore}[1]{}

\newcommand{\myparagraph}[1]{{\smallskip\noindent\textbf{#1}}}
\newcommand{\emparagraph}[1]{{\smallskip\noindent\textit{#1}}}

\newcommand{\calA}{{\cal A}}
\newcommand{\calB}{{\cal B}}

\newcommand{\tildeC}{{\tilde{C}}}  
\newcommand{\tildeN}{{\tilde{N}}}

\newcommand{\tildeO}{{\tilde{O}}}
\newcommand{\tildeOmega}{{\tilde{\Omega}}}

\newcommand{\braced}[1]{{ \left\{ #1 \right\} }}

\newcommand{\brackd}[1]{{ \left[ #1 \right] }}

\newcommand{\suchthat}{{\;:\;}}

\newcommand{\RoundRobin}{{\textsc{RoundRobin}}}
\newcommand{\rws}{{\textit{rws}}}
\newcommand{\algAcyclicGather}{\mbox{\textsc{AcyGather}}}  
\newcommand{\algAcyclicGatherWithAck}{\mbox{\textsc{AcyGatherAck}}} 
\newcommand{\algArbitraryGather}{\mbox{\textsc{ArbGather}}}
\newcommand{\algSccGossip}{{\textsc{SccGossip}}}

\newcommand{\Selector}[1]{{\textsc{$#1$-Select}}}
\newcommand{\HalfSelector}[1]{{\textsc{$#1$-HalfSelect}}}

\newcommand{\ACY}{{\textsf{ACY}}}
\newcommand{\SCC}{{\textsf{SCC}}}

\newcommand{\ingraphinG}{G^-}
\newcommand{\sccomp}{{C}}  
\newcommand{\tildesccomp}{{\tildeC}}  
\newcommand{\inneighb}{{N^{-}}}
\newcommand{\sccinneighb}{{N^{-}_{\scriptscriptstyle\textrm{scc}}}}
\newcommand{\acyinneighb}{{N^{-}_{\scriptscriptstyle\textrm{acy}}}}
\newcommand{\tildeacginneighb}{{{\tildeN}^{-}_{\scriptscriptstyle\textrm{acy}}}} 
\newcommand{\acttime}{\alpha}
\newcommand{\sccacttime}{\alpha_{\scriptscriptstyle\textrm{scc}}}
\newcommand{\acyacttime}{\alpha_{\scriptscriptstyle\textrm{acy}}}

\newcommand{\sccruntime}{T_{\scriptscriptstyle\textrm{SCC}}}

\newcommand{\maxpath}{{\delta}}
 \newcommand{\maxmaxpath}{{\delta^\ast}}

\newtheorem{claim}{Claim}
\newtheorem{observation}{Observation}

\newenvironment{bigeqn*}{\large\begin{eqnarray*}}{\end{eqnarray*}}

\newcommand{\half}{{\textstyle\frac{1}{2}}}

\begin{document}
	
\maketitle

\begin{abstract}
	In the ad-hoc radio network model, nodes communicate with their neighbors
    via radio signals, without knowing the topology of the underlying digraph.
	We study the information gathering problem, where each node has a piece of
	information called a \emph{rumor}, and the objective is to transmit all
	rumors to a designated target node.  For the model without any collision detection
	we provide an $\tildeO(n^{1.5})$ deteministic protocol, 
	significantly improving the trivial bound of $O(n^2)$.  
	We also consider a model with a mild form of collision detection, 
	where a node receives a 1-bit acknowledgement if its transmission was received by at least one
	out-neighbor. For this model we give a $\tildeO(n)$ deterministic
	protocol for information gathering in acyclic graphs.
\end{abstract}


\section{Introduction}
\label{sec: introduction}

 
We address the problem of information gathering in ad-hoc radio networks. A \emph{radio network}
is represented by a directed graph (digraph) $G$, whose nodes represent radio transmitters/receivers and
directed edges represent their transmission ranges; that is, an edge $(u,v)$ is present in the digraph if
and only if node $v$ is within the range of node $u$. 
When a node $u$ transmits a message, this message is immediately sent out to all its out-neighbors. 
However, a message may be prevented from reaching some out-neighbors of $u$ if it collides with
messages from other nodes. A collision occurs at a node $v$ if two or more in-neighbors of $v$ transmit at the
same time, in which case $v$ will not receive any of their messages, and it will not even know that they transmitted.

Radio networks, as defined above, constitute a useful abstract model for studying protocols for
information dissemination in networks where communication is achieved via broadcast channels, 
as opposed to one-to-one links. Such networks do not need to
necessarily utilize radio technology; for example, in local area networks based on  the
ethernet protocol all nodes communicate by broadcasting information through a shared carrier.
Different variants of this model have been considered in the literature, depending
on the assumptions about the node labels (that is, identifiers), on the knowledge of the underlying topology, 
and on allowed message size.  In this work we assume that nodes are labelled $0,1,...,n-1$,
where $n$ is the network size. (All our results remain valid if the labels are selected from the range $0,1,...,O(n)$.)
We focus on the \emph{ad-hoc model}, where the digraph's topology is
uknown when the computation starts, and a protocol needs to complete its task within a desired time bound,
no matter what the topology is. At the beginning of the computation each node $v$ is in possession
of a unique piece of information, that we refer to as a \emph{rumor}. Different
communication primitives are defined by specifying how these rumors need to be disseminated across the network. In this paper
we do not make any assumptions about the size of transmitted messages; thus a node can aggregate multiple
rumors and transmit them in one message. In fact, it could as well transmit as one message the complete history of its past computation.

Two most studied information dissemination primitives for this model are
broadcasting and gossiping. In \emph{broadcasting} (or \emph{one-to-all communication}), 
a single source node $s$ attempts to deliver its
rumor to all nodes in the network. For broadcasting to be meaningful, we need to assume that all
nodes in $G$ are reachable from $s$. In \emph{gossiping} (or \emph{all-to-all communication}),
the objective is to distribute all rumors to all nodes in the network, under the assumption that $G$ is strongly connected. 
Both these primitives can be solved in time $O(n^2)$ by a simple protocol called {\RoundRobin} where all
nodes transmit cyclically one at a time (see Section~\ref{sec: preliminaries}). 
Past research on ad-hoc radio networks focussed on designing protocols that improve this trivial bound.

For broadcasting, gradual improvements in the running time have been reported since early 2000's
~\cite{Chrobak_etal_fast_02,Kowalski_Pelc_faster_04,Bruschi_etal_lower_bound_97,Chlebus_etal_broadcast_02,DeMarco_08,Czumaj_Rytter_broadcast_06},
culminating in the upper bound of $O(n\log D\log\log (D\Delta/n))$ in~\cite{Czumaj_Davies_faster_16},
where $D$ denotes the diameter of $G$ and $\Delta$ its maximum in-degree.
This is already almost tight, as the lower bound of $\Omega(n\log D)$ is known~\cite{Clementi_etal_distributed_02}.
For randomized algorithms, the gap between lower and upper bounds is also
almost completely closed, see~\cite{alon_etal_lower_bound_91,Kushilevitz_Mansour_lower_bound_98,Czumaj_Rytter_broadcast_06}.

In case of gossiping, major open problems remain. The upper bound of $O(n^2)$ was improved
to $\tildeO(n^{1.5})$ in~\cite{Chrobak_etal_fast_02,Xu_det_gossip_03} and then later to
$\tildeO(n^{4/3})$ in~\cite{Gasieniec_etal_det_gossip_04}, and no better bound is currently known\footnote{We use notation
$\tildeO(f(n))$ to conceal poly-logarithmic factors; that is, $g(n) = \tildeO(f(n))$ iff $g(n) = O(f(n)\log^c n)$ for some
constant $c$. Also, we write $f(n) = \tildeOmega(g(n))$ if and only if $g(n) = \tildeO(f(n))$.}. 
No lower bound better than $\Omega(n\log n)$ (that follows from~\cite{Clementi_etal_distributed_02}) is known. 
In contrast, in the randomized case it is possible to achieve gossiping in time
$\tildeO(n)$~\cite{Czumaj_Rytter_broadcast_06,Liu_Prabhakaran_randomized_02,Chrobak_etal_randomized_04}.

The reader is referred to survey  papers~\cite{Gasieniec_survey_gossiping_09,Jurdzinski_K_enc_wakeup_2015,Gasieniec_deterministic_broadcasting_2016,Peleg_broadcasting_review_2007,Itai_randomized_broadcasting_encyclopedia_2016} that contain more information about information dissemination protocols in different variants of radio networks.

In this paper we address the problem of \emph{information gathering} (that is, \emph{all-to-one communication}). 
In this problem, similar to gossiping, each node $v$ has its own rumor, and the objective is to deliver these rumors to a
designated target node $t$. (We assume that $t$ is reachable from all nodes in $G$.)   

The problem of information gathering for trees was introduced in~\cite{ChrobakCGK_tree_gather_18}, where
an $O(n)$-time algorithm was presented. Other results in~\cite{ChrobakCGK_tree_gather_18} include 
algorithms for the model without rumor aggregation or the model with transmission acknowledgements.


\myparagraph{Our results.}
Our main result, in Section~\ref{sec: tildeO(n1.5) protocol arbitrary}, 
is a deterministic protocol that solves the information gathering problem in arbitrary ad-hoc networks in time $\tildeO(n^{1.5})$.
To our knowledge this is the first such a protocol that achieves running time faster than the trivial $O(n^2)$ bound.
One of our key technical contributions is in solving this problem in time $\tildeO(n^{1.5})$ for acyclic 
graphs (Section~\ref{sec: tildeO(n1.5) protocol acyclic})
where any protocols developed earlier for gossiping, that rely on feedback (see the discussion below), are not applicable. 
This algorithm for acyclic graphs is based on careful application of combinatorial structures called strong selectors,
combined with a novel amortization technique to measure progress of the algorithm. To extend this protocol to arbitrary graphs, 
we integrate it with a gossiping protocol. Roughly, the two sub-protocols run in parallel, with
the sub-protocol for acyclic graphs transferring information between strongly connected components, while the gossiping 
sub-protocol disseminates it within each strongly connected component. This requires overcoming two challenges.
One is that the partition of $G$ into strongly connected components is not actually known, 
so the combined protocol needs to gradually ``learn'' the connectivity structure of $G$ while it executes.  
The second challenges is in synchronizing the computation of the two sub-protocols, since they are based on
entirely different principles.

In the second part of the paper, in Section~\ref{sec: protocol with acknowledgement}, we
consider a slight relaxation of our model by allowing a mild form of collision detection.
In this new model each node $v$, after each transmission, receives a 1-bit acknowledgement indicating 
whether its transmission was received by at least one out-neighbor.  With this assumption, we provide an $\tildeO(n)$-time 
algorithm for information dissemination in acyclic radio networks.


\myparagraph{Additional context and motivations.}
If $G$ is strongly connected then information gathering and gossiping are equivalent. Trivially, 
a gossiping algorithm gathers all rumors in $t$, solving the information gathering problem.
On the other hand, one can solve the gossiping problem by running an information gathering protocol
followed by any $\tildeO(n)$-time broadcasting protocol with source node $t$.  Thus, counter-intuitively,
information gathering can be thought of as \emph{an extension} of gossiping, since it applies to a broader class of graphs.

The crucial challenge in designing protocols for information gathering is \emph{lack of feedback},
namely that the nodes in the network do not receive any information about the fate of their
transmissions.  This should be contrasted with the gossiping problem where, due to the 
assumption of strong connectivity, a node can eventually learn whether its earlier transmissions were
successful. In fact, the existing protocols for gossiping critically rely on this
feature, as they use it to identify nodes that have collected a large number of rumors, and
then they broadcast these rumors to the whole network, thus removing them from consideration and reducing congestion.  

Some evidence that feedback might help to speed up information gathering can be found in~\cite{chrobak_costello_faster_18},
where the authors developed an $O(n)$-time protocol for trees if nodes receive (immediate) acknowledgements of
successful transmissions, while the best known upper bound for this problem without feedback is $O(n\log\log n)$.

Various forms of feedback have been studied in the past in the context of \emph{contention resolution} 
for multiple-access channels (MAC), where nodes communicate via a single shared
challel. (Ethernet is one example.) Depending on more specific characteristics of
this shared channel, one can model this problem as the information gathering  problem either on a complete
graph or a star graph, which is a collection of $n$ nodes connected by directed edges to the target node $t$. (See
\cite{DeMarco_Kowalski_13,demarco_kowalski_conflict_15,Anta_etal_13} for information about contention resolution protocols.) 
For instance, in~\cite{ChrobakCGK_tree_gather_18} a tight bound of $\Theta(n\log n)$ was given
for randomized information gathering on star graphs (or MACs) even if the nodes have no labels (are indistinguishable)
and receive no feedback.       

As explained earlier, in our model rumor aggregation is allowed.
This capability is needed to beat the $O(n^2)$ upper bound, as
without rumor aggregation it is quite easy to show a lower bound of $\Omega(n^2)$ for
both gossiping and information gathering, 
and even for randomized algorithms and with the topology known~\cite{GasieniecP_gossip_unit_02}.

Interestingly, if we allow randomization, the randomized gossiping algorithms 
in~\cite{Chrobak_etal_randomized_04,Liu_Prabhakaran_randomized_02}
can be adapted to information gathering without increasing the running time. Thus
randomization can not only help to overcome collisions, but also lack of feedback.


\section{Preliminaries}
\label{sec: preliminaries}




\myparagraph{Graph terminology.}
Throughout the paper, we assume that the radio network is represented by a digraph (directed graph) $G$
with a distinguished target node $t$ that is reachable from all other nodes. By $n = |G|$ 
we denote the number of nodes in $G$.
We will treat $G$ both as a set of vertices and edges, and write $u\in G$ if $u$ is a node of $G$ and
$(u,v)\in G$ if $(u,v)$ is an edge of $G$.
If $(u,v)\in G$ then we refer to $u$ as the \emph{in-neighbor} of $v$ and to $v$ as the \emph{out-neighbor} of $u$.
For any node $v$, by $\inneighb(v) = \braced{u\in G\suchthat (u,v)\in G}$ we denote the set of its in-neighbors.

For brevity, we will refer to strongly connected components of $G$ as \emph{sc-components}.
For each node $v$, the sc-component containing $v$ will be denoted by $\sccomp(v)$.
We partition the set of in-neighbors of $v$ into those that belong to $\sccomp(v)$ and those that do not:
$\sccinneighb(v) = \inneighb(v) \cap \sccomp(v)$ and $\acyinneighb(v) =\inneighb(v) \setminus \sccomp(v)$.

The \emph{in-graph} of $v$ in $G$, denoted $\ingraphinG(v)$, is the set of all nodes of $G$ from which $v$ is reachable
(via a directed path). We extend this definition in a natural way to sc-components of $G$; if $A$ is an sc-component
then its in-graph is $\ingraphinG(A) = \bigcup_{v\in A} \ingraphinG(v)$.


\myparagraph{Radio networks.}
As mentioned in the introduction we assume that each node of $G$ has a unique label from
the set $\brackd{n} = \braced{0,1,...,n-1}$. For convenience, we will identify nodes
with their labels, so a ``node $u$'' really means the node with label $u$.

The time is divided into discrete \emph{time steps} numbered with non-negative integers.
We assume that all nodes start to execute the protocol simultaneously at time step $0$.
In the formal model of radio networks, at each step each node can be either in a \emph{transmitting state},
when it can transmit a message, or \emph{receiving state}, when it can only
listen to transmissions from other nodes. We will show below, however, that we can relax these
restrictions and allow a node to simultaneously listen and transmit at each step.
Only one message can be transmitted at each step. This is not an essential restriction because,
as already mentioned, we are not imposing any restrictions on the size or format of messages transmitted
by nodes. However, a message transmited at a given step cannot depend on the message (if any) 
received in the same step.

If a node $u$ transmits a message at a time $\tau$, this message reaches all out-neighbors of $u$
in the same step. If $v$ is one of these out-neighbors, and if $u$ is the only in-neighor of $v$
that transmits at time $\tau$, then $v$ will receive this message. However, if there are two
or more in-neighbors of $v$ that transmit at time $\tau$ then a \emph{collision occurs}, and
$v$ does not receive any information. In other words, collisions are indistinguishable from absence
of transmissions. There is no feedback mechanism available in this model, that is a sender of a
message does not receive any information as to whether its transmission was successful or not.
(We will relax this restriction later in Section~\ref{sec: protocol with acknowledgement}.)


\myparagraph{Selectors.}
A \emph{strong $(n,k)$-selector} is a sequence of label sets
$(S_0,S_1,...,S_{\ell-1})$ (that is, $S_i\subseteq[n]$ for each $i$)
that ``singles out'' each label from each subset of at most $k$ labels, in the following sense:
for each $X\subseteq \brackd{n}$  with $|X|\le k$ and each $x\in X$ there
is an index $i$ such that $S_i\cap X = \braced{x}$. It is known~\cite{Clementi_etal_01} that
there exist strong $(n,k)$-selectors of size $\ell = O(k^2\log n)$.

Such selectors are often used for designing protocols for ad-hoc radio networks.
The intuition is this: Consider a protocol that cyclically ``runs'' a strong $(n,k)$-selector;
that is, each node $w$ transmits in a step $\tau$ if and only if $w \in S_{\tau\bmod\ell}$.
Suppose that $u$ starts transmitting its message at some time step and then follows this protocol.
If $v$ is an out-neighbor of $u$ and $v$'s in-degree is at most $k$,
then $v$ will successfully receive $u$'s message in at most $ O(k^2\log n)$ steps,
independently of the label assignment.
Another basic protocol that is often used is called {\RoundRobin}. In this protocol
all nodes transmit cyclically one by one; that is
each node  $w$ transmits in a step $\tau$ if and only if $w = \tau\bmod n$.
In {\RoundRobin} there are no collisions so, in the setting above,
node $u$ will successfully transmit its message to $v$ in at most $n$ time steps.
Note that a protocol based on a strong $(n,k)$-selector can be faster than {\RoundRobin} only
when $k = O(\sqrt{n/\log n})$.

For all $j = 0,1,..., \half\log n$, by
$\Selector{2^j} = (S^j_0,S^j_1,...,S^j_{\ell_j-1})$ we will denote a strong $(n,2^j)$-selector
of size $\ell_j =O(4^j\log n)$. Without loss of generality we can assume that
$\ell_{j+1} = 4\ell_j$ for all $j \le \half\log n-1$.

\emph{Note:} To avoid clutter, in the paragraph above, as well as later throughout the paper,
we omit the notation for rounding
and assume that in all formulas representing integer quantities (the number of nodes, steps,
etc.) their values are appropriately rounded. This will not affect asymptotic running time estimates.

In Section~\ref{sec: protocol with acknowledgement}, where we consider transmissions with acknowledgements, 
it will be desirable to have many (but not necessarily all) of a collection of competing 
in-neighbors of a node transmit successfully. For this purpose we will there introduce a different type
of selectors.


\myparagraph{Simplifying assumptions.}
To streamline the description of our algorithms, in the paper we will assume a relaxed communication model with two
additional features:
\begin{description}
	\item{(MFC)} 
			We assume that some number $\kappa$ of radio \emph{frequency channels}, numbered $0,1,...,\kappa-1$,
			is available for communication. In a single step, a node can use all frequencies simultaneously.  
	\item{(SRT)}
			Further, for each frequency $f$, a node can receive and transmit at frequency $f$ in a single step.
			The restriction is that the messages transmitted at all frequencies in any step do not depend on
			the messages received in this step.
\end{description}
Below we explain how this relaxed model can be simulated using the standard radio network model,
increasing the running time by factor $O(\kappa)$; that is, any protocol that uses 
features (MFC) and (SRT) and runs in time $O(T)$ can be converted into a protocol in
the standard model whose running time is $O(\kappa T)$.  Since $\kappa = O(\log n)$ in our
protocols, their $\tildeO(\cdot)$-complexity is not affected.


\emparagraph{Simulating multiple frequencies.}
We first explain how we can convert a protocol $\calA$ that uses $\kappa$ frequencies and runs in time $O(T)$
into a protocol $\calA'$ that uses only one frequency and runs in time $O(\kappa T$).
This can be done by straightforward time multiplexing. In more detail: $\calA'$
organizes all time steps $0,1,2,...$ into \emph{rounds}.
Each round $r = 0,1,2,...$ consists of $\kappa$ consecutive steps $r \kappa, r \kappa+1, ..., r\kappa+\kappa-1$.
Each step $s$ of $\calA$ is simulated by round $s$ of $\calA'$. For each frequency $f$, the message transmitted
at frequency $f$ by $\calA$ is transmitted by $\calA'$ in step $s\kappa+f$, that is the
$f$th step of round $s$. At the end of
round $s$, $\calA'$ will know all messages received in this round, so it will know what
messages would $\calA$ receive in step $s$, and therefore it knows the state of $\calA$ and
can determine the transmissions of $A$ in the next step.


\begin{figure}[ht]
\begin{center}
\includegraphics[width = 5.5in]{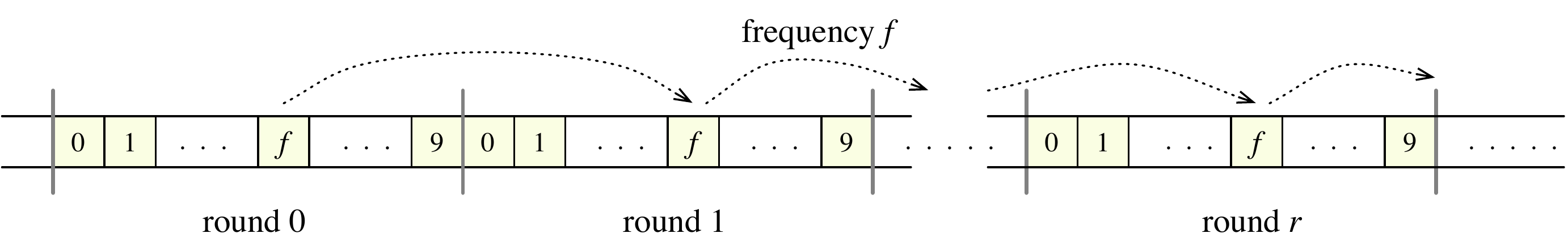}
\end{center}
\caption{Partition of $\calA'$'s time steps into rounds, for $\kappa = 10$ frequencies.}
\label{fig: simulation of multiple frequencies}
\end{figure}



\emparagraph{Simulating simultaneous receiving/transmitting.}
By the argument above, we can assume that we have only one frequency channel.
We claim that we can disallow simultaneous receiving and transmitting
at the cost of only adding a logarithmic factor to the running time.
To see this, suppose that $\calB$ is some transmission
protocol where nodes can transmit and listen at the same time.  (Recall that the
transmission of $\calB$ at any step does not depend on the information it receives
in the same step.) We use a strong $(n,2)$-selector $\Selector{2} = (S^1_i)_i$ of size $\ell_1 = \tildeO(1)$.
We replace each step $\tau$ of $\calB$ by a time segment $I_\tau$ of length $\ell_1$.
For any node $u$ and any $i = 0,1,...,|I_\tau|-1$,  if $v\in S^1_i$ then at the $i$th step of segment $I_\tau$ node
$v$ transmits whatever message it would transmit in $\calB$ at time $\tau$;
otherwise $v$ is in the receiving state. By definition, in this new protocol $\calB'$ 
nodes do not transmit and receive at the same time. Further, for any edge $(u,v)$, 
if $u$ transmitted successfully to $v$ in step $\tau$ of $\calB$, in $\calB'$ there will be a time step
within $I_\tau$ at which $u$ is in the transmitting state and $v$ is in the receiving
state, guaranteeing that $u$'s message will reach $v$.

In fact, for the type of protocols presented in the paper, allowing simultaneous reception and transmisison
does not affect the asymptotic running time at all. Our protocols are based on
strong selectors and {\RoundRobin}. In case of {\RoundRobin}, the simultaneous reception and transmisison
capability is (trivially) not needed. For selector-based protocols, the argument how this
capability can be removed was given in~\cite{chrobak_costello_faster_18}.
Roughly, the idea is that whenever a protocol uses a strong $(n,k)$-selector, this
selector can be replaced by a strong $(n,k+1)$-selector (whose size is asymptotically the same).
This guarantees that, during each complete cycle (of length $O(k^2\log n)$) of this selector, 
for any node $v$ with $k$ in-neighbors and any $v$'s in-neighbor $u$
there will be a step when $v$ is in the receiving state and $u$ is the only in-neighbor in the transmitting state.


\section{$\tildeO(n^{1.5})$-Time Protocol for Acyclic Digraphs}
\label{sec: tildeO(n1.5) protocol acyclic}



We first consider ad-hoc radio networks whose underlying digraph $G$ is acyclic and has one
designated target node $t$ that is reachable from all other nodes in $G$. 
We give a deterministic information gathering protocol that gathers all rumors in
the target node $t$ in time $\tildeO(n^{1.5})$, independently of the topology of $G$.

In the algorithm we will assume that each vertex knows the labels of its in-neighbors. This can be easily
achieved in time $O(n)$ by pre-processing that consists of one cycle of {\RoundRobin}, where each 
node transmits only its own label. As explained in Section~\ref{sec: preliminaries},  
we also make Assumptions~(MFC) and (SRT), namely that the protocol has multiple frequency channels available and
on each frequency it can simultaneously receive and transmit messages at each step.

Let $\theta = \half(\log n - \log\log n)+2$.
In the algorithm below we use a sequence of $\theta+1$ values  $\beta_0,\beta_1,...,\beta_\theta$,
defined as follows: $\beta_{0} = 0$, $\beta_{j} = \sum_{g < j} \ell_g$ for $j = 1,...,\theta-1$,
and $\beta_{\theta} =  \sum_{g < \theta} \ell_g + n$. 


\myparagraph{Protocol~{\algAcyclicGather}.}
The algorithm uses $\theta$ frequencies numbered $0,1,...,\theta-1$.
The intuition is that each frequency $j \le \theta-2$ will be used to run selector $\Selector{2^j}$,
while frequency $\theta-1$ will be used to run $\RoundRobin$.

At each step, a node could be \emph{dormant} or \emph{active}. Dormant nodes do not
transmit; active nodes may or may not transmit. A node $v$ is active during its
\emph{activity period} $[\acttime(v),\acttime(v) + \beta_{\theta})$, where $\acttime(v)$ is
referred to as the \emph{activation step of $v$}, and is defined below.

If $v$ is a source node (that is, its in-degree is $0$), then $\acttime(v) = 0$. Otherwise
$\acttime(v)$ is determined by the messages received by $v$, as follows.
Each message transmitted by a node $u$ contains the following information:
(i) all rumors collected by $u$, including its own,
(ii) the label of $u$, and 
(iii) another value called  \emph{recommended wake-up step} and denoted $\rws_u$,
to be defined shortly. 
For a non-source node $v$ and its in-neighbor $u$, denote by $\rws^1_{u,v}$ the
\emph{first} $\rws_u$ value received by $v$ from $u$. (This may not be the first
$\rws_u$ value \emph{transmitted} by $u$, since earlier transmissions of $u$ might have collided at $v$.)
Node $v$ waits until it receives messages from all its in-neighbors, and, as soon as this happens,
if $u$ is the last in-neighbor of $v$ that successfully transmitted to $v$, then $v$ sets $\acttime(v) = \rws^1_{u,v}$.
(Occasionally we will write $\rws^1(u,v)$ instead of $\rws^1_{u,v}$, to avoid multi-level indexing.) 

\begin{figure}[ht]
\begin{center}
\includegraphics[width = 4.5in]{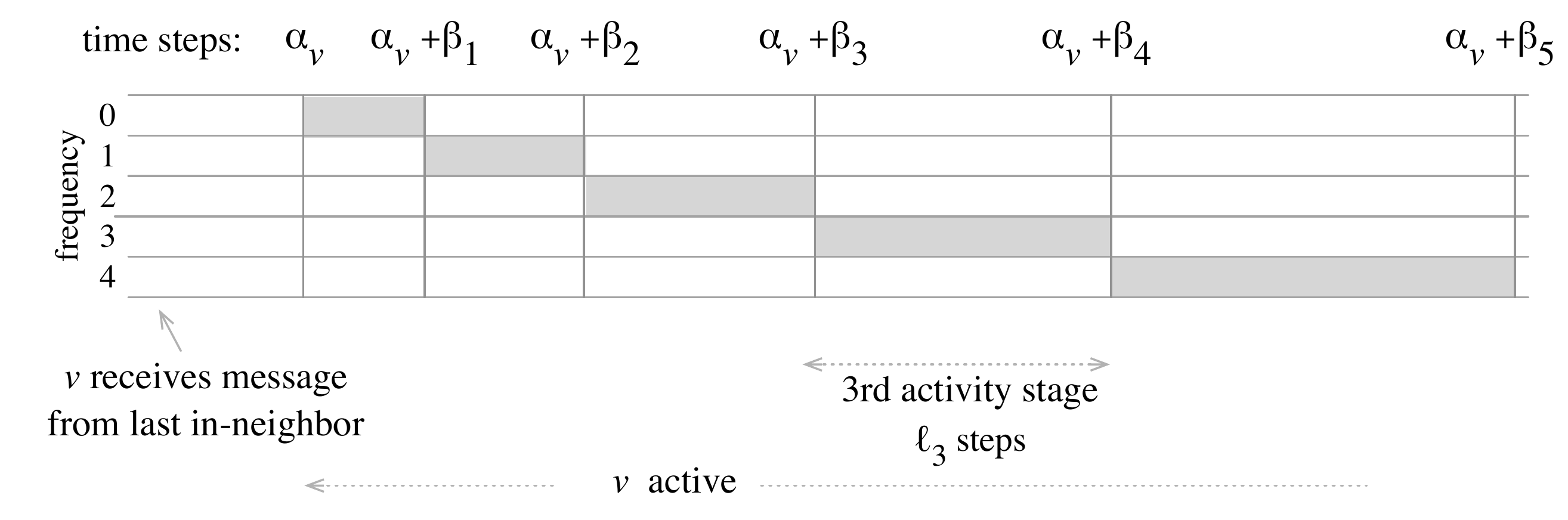}
\end{center}   
\caption{Illustration of activity stages. (The picture is not up to scale. 
In reality the length of activity stages increases at rate $4$.) Shaded regions
show frequencies used in different activity stages.} 
\label{fig: activity stages}
\end{figure}

The activity period $[\acttime(v),\acttime(v) + \beta_{\theta})$ of $v$ is divided into $\theta$ \emph{activity stages}, where,
for $j = 0,1,...,\theta-1$, the $j$th activity stage  consists of the time interval
 $[\acttime(v)+\beta_{j},\acttime(v)+\beta_{j+1})$. (See Figure~\ref{fig: activity stages}.)   
During its $j$th activity stage, for $j \le \theta-2$, node $v$ transmits according to selector
$\Selector{2^j}$ using frequency $j$. During the $(\theta-1)$th activity stage,
the protocol transmits using $\RoundRobin$ on frequency $\theta-1$.
The recommended wake-up step value included in $v$'s messages during its $j$th activity stage is 
$\rws_{v} = \acttime(v) + \beta_{j+1}$. At all other times $v$ does not transmit.


\myparagraph{Correctness.} 
We first note that the algorithm is correct, in the sense that each rumor will eventually reach the target
node $t$. This is true because once a node becomes active, it is guaranteed to successfully
transmit its message to its all out-neighbors using the {\RoundRobin} protocol during its last activity stage.


\myparagraph{Running time.}
Next, we show that Protocol~{\algAcyclicGather} completes information gathering in time $\tildeO(n^{1.5})$. 
To establish this bound, we choose in the graph $G$ a \emph{critical path} $P =  (v_0, v_1,...,v_p = t)$, 
defined as follows: for each $a = p-1,p-2,...,0$, $v_a$ is the in-neighbor of $v_{a+1}$
who was last to successfully transmit to $v_{a+1}$  (thus $\acttime(v_{a+1}) = \rws^1_{v_a,v_{a+1}}$), and $v_0$ is a source node. 
(Note that, since we define this path in the backwards order, the indexing of
the nodes $v_a$ can be determined only after we determine the whole path).  
The overall running time is upper-bounded by the time for the rumor of $v_0$ to reach $t$ along $P$.   


\begin{figure}[ht] 
\begin{center}
\includegraphics[width = 3.25in]{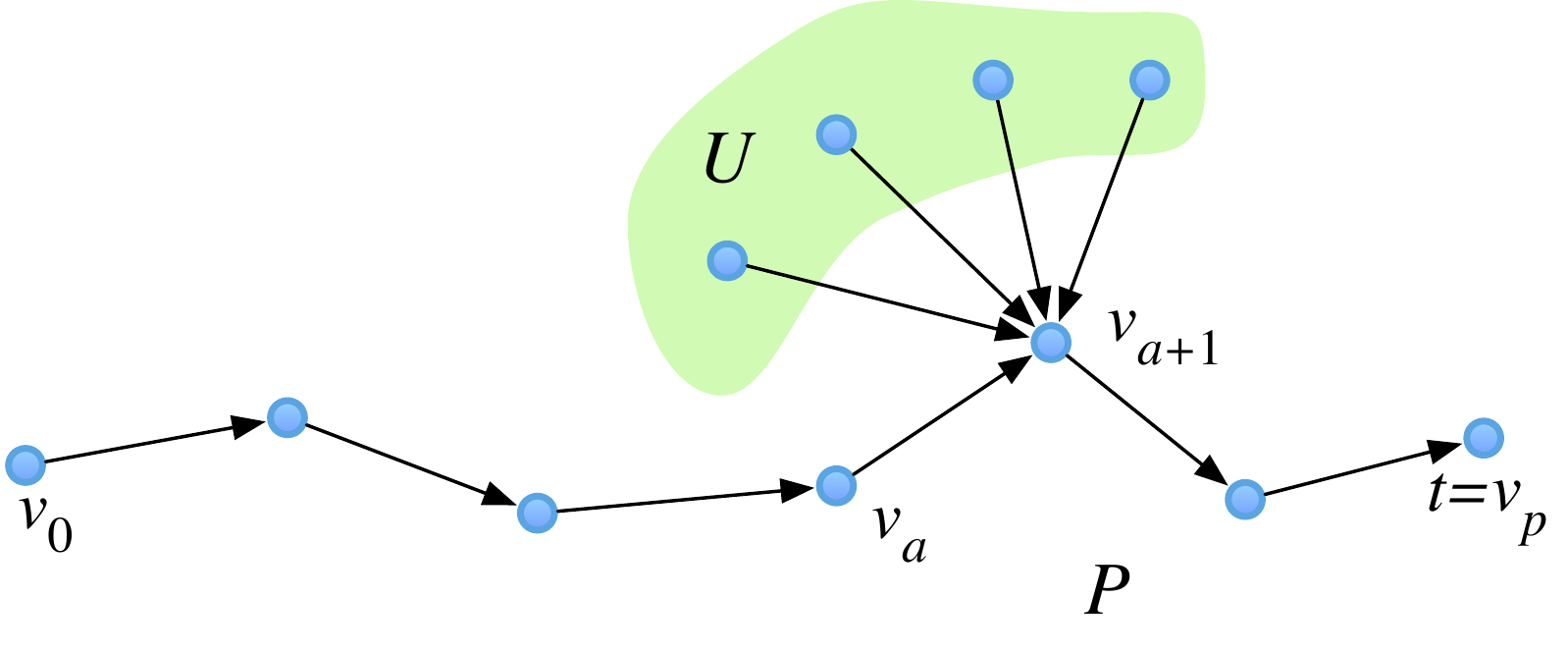} 
\end{center}   
\caption{Illustration of the time analysis for acyclic graphs.} 
\label{fig: time analysis acyclic}     
\end{figure}


If at a step $\tau$ a node $v$ is in its $j$-th activity stage (that is, 
$\tau \in [\acttime(v)+\beta_j, \acttime(v)+\beta_{j+1})$) then we refer to $j$ 
as \emph{$v$'s stage index in step $\tau$}. We extend this (artificially) to dormant nodes as follows:
if $v$ has not yet started its activity period then its stage index is $-1$, and
if $v$ has already completed its activity period then its stage index is $\theta$.
The stage index of each node is incremented $\theta+1 = O(\log n)$ times, so the
total number of these increments, over all nodes and over the whole computation, is $O(n\log n) = \tildeO(n)$.

Now consider some node $v_a$ on $P$. (See Figure~\ref{fig: time analysis acyclic}.)  Our argument is
based on the following key lemma.

\begin{lemma}\label{lem: stage increments} 
There are $\tildeOmega(n^{-1/2} (\, \acttime(v_{a+1}) - \acttime(v_a)\,))$ 
stage index increments in the time interval $[ \acttime(v_a), \acttime(v_{a+1}) )$.
\end{lemma}

Before we prove Lemma~\ref{lem: stage increments}, we argue that this lemma
is sufficient to establish our $\tildeO(n^{1.5})$ upper bound. 
Let $T$ be the running time of  Protocol~{\algAcyclicGather}.
Since $\acttime(v_0) = 0$ and $T \le \acttime(v_p)$,  we can bound the running time as 
$T \le \sum_{a=0}^{p-1} ( \acttime(v_{a+1}) - \acttime(v_a) )$. Then Lemma~\ref{lem: stage increments}
implies that the total number of stage index increments during the computation
is $\tildeOmega(n^{-1/2}T)$.   Since this number is also $\tildeO(n)$, it gives us that $T = \tildeO(n^{1.5})$.


\begin{proof}
We now prove Lemma~\ref{lem: stage increments}. Suppose that
$v_a$ succeeds first time in transmitting its message  to $v_{a+1}$ during its $h$-th activity stage. 

\begin{claim}\label{clam: difference of alphas} 
For $a<p$ and $h < \theta-1$ we have $\acttime(v_{a+1}) - \acttime(v_{a}) =  \tildeO(4^h)$.
\end{claim}

This claim follows from the definition of $P$, as $\acttime(v_{a+1}) = \rws^1_{v_a,v_{a+1}} = \acttime(v_{a}) + \beta_{h+1}$,
and $\beta_{h+1} =  \sum_{g < h} \ell_g = O(4^h\log n)$.

We now consider three cases, depending on the value of $h$.
First, if $h=0$, then there is at least one stage increment in $[ \acttime(v_a), \acttime(v_{a+1}) )$
(namely the increment of the stage index of $v_a$ from $-1$ to $0$) and
$\acttime(v_{a+1}) - \acttime(v_a) = \ell_0 = O(\log n)$, so the lemma holds trivially.

Next, suppose that $1\le h \le \theta-2$. By the choice of $h$, $v_a$ has not succeeded in its $(h-1)$th activity stage 
$[\acttime(v_a) + \beta_{h-1},\acttime(v_a) + \beta_{h})$. 
Let $U$ be the set of in-neighbors of $v_{a+1}$ (including $v_a$) whose $(h-1)$th activity stage overlapped that of $v_a$. 

\begin{claim}\label{cla: size of U} 
$|U| > 2^{h-1}$.
\end{claim}

To justify Claim~\ref{cla: size of U}, we argue by contradiction. Suppose that $|U| \le 2^{h-1}$.
During this activity stage $v_a$ transmitted according to $\Selector{2^{h-1}}$ using only frequency $h-1$.
Further, by the definition of the protocol, at each step of this stage the 
in-neighbors of $v_{a+1}$ with stage index other than $h-1$ did not use frequency $h-1$ for transmissions.
So the transmissions from $v_a$ to $v_{a+1}$ in this stage can only conflict with transmissions from 
$U\setminus\braced{v_a}$ to $v_{a+1}$.
The definition of strong selectors and the assumption that $|U|\le 2^{h-1}$  imply that then $v_a$ would have successfully
transmitted to $v_{a+1}$ during its $(h-1)$th activity stage, contradicting the definition of $h$.
Thus Claim~\ref{cla: size of U} is indeed true.

The $(h-1)$th activity stage lasts $\ell_{h-1}$ steps so all the $(h-1)$th activity stages of the nodes in $U$ end
before time $\acttime(v_a) + \beta_h + \ell_{h-1}  < \acttime(v_a) + \beta_{h+1} = \acttime(v_{a+1})$.                    
This implies that in the interval $[ \acttime(v_a),\acttime(v_{a+1}) )$ the number of stage index increments is at least 
\begin{equation*}
|U| \;\ge\; 2^{h-1} \;=\; \half \cdot 2^{-h}\cdot 4^h 
				\;=\; \tildeOmega(n^{-1/2} ( \acttime(v_{a+1}) - \acttime(v_a) ) ),
\end{equation*}
because $h\le \half\log n$ and $\acttime( v_{a+1} ) - \acttime(v_a) = \tildeO(4^h)$.  
This completes the proof of the lemma when $1\le h \le \theta-2$.

Finally, consider the case when $h = \theta-1$. Then $\acttime(v_{a+1}) - \acttime(v_a) = n$.  But, by the choice of $h$,
$v_a$ has not succeeded in its $(h-1)$th activity stage, where $h-1 = \half (\log n-\log\log n)$.
A similar argument as above gives us that the number of stage index increments during 
$v_a$'s $(h-1)$th activity stage is $\tildeOmega(n^{1/2})$, implying Lemma~\ref{lem: stage increments}.
\end{proof}


\myparagraph{More precise time bound.}
We have established that Algorithm~{\algAcyclicGather} runs in time $\tildeO(n^{1.5})$
on acyclic graphs.
For a more precise bound, let us now determine the exponent of the logarithmic factor in this bound:
one factor $O(\log n)$ is needed to simulate multiple frequencies with one,
one factor $O(\log n)$ appears in the bound for the length of selectors, and
we have another factor  $O(\log n)$ that we ignored in the amortized analysis,
since the number of stage index increments is $O(n\log n)$ (while we 
used the bound of $\tildeO(n)$). This gives us the main result of this
section:


\begin{theorem}\label{lem: running time of  AcyclicGather}
Let $G$ be an acyclic directed graph with $n$ vertices and a designated target node reachable from
all other nodes.	
Algorithm~{\algAcyclicGather} completes information gathering on $G$ in time $O(n^{1.5}\log^3 n)$.
\end{theorem}


\section{$\tildeO(n^{1.5})$-Time Protocol for Arbitrary Digraphs}
\label{sec: tildeO(n1.5) protocol arbitrary}
  


We now extend our information gathering protocol~{\algAcyclicGather} from 
Section~\ref{sec: tildeO(n1.5) protocol acyclic} to arbitrary digraphs, retaining running time $\tildeO(n^{1.5})$. 
Throughout this section $G$ will denote an $n$-vertex digraph with a designated target node $t$
that is reachable from all other nodes in $G$.

The main obstacle we need to overcome is that protocol~{\algAcyclicGather} critically depends on
on $G$ being acyclic.  For instance, in that protocol each node waits
until it receives messages from all its in-neighbors. If cycles are present in $G$, this leads to a deadlock, where each
node in a cycle waits for its predecessor. On the other hand, the known gossiping 
protocols~\cite{Chrobak_etal_fast_02,Xu_det_gossip_03,Gasieniec_etal_det_gossip_04}  do not work correctly if the graph
is not strongly connected, because they rely on broadcasting to periodically flush out some rumors from the
system and on leader election to synchronize computation. 

The idea behind our solution is
to integrate protocol~{\algAcyclicGather} with the gossiping protocol from~\cite{Gasieniec_etal_det_gossip_04},
using {\algAcyclicGather} to transmit information between different sc-components of $G$
and using gossiping to disseminate information within sc-components. The idea is natural but it faces several
technical challenges. One challenge is that the sc-components of $G$ are actually not known. In fact, a node
$v$ doesn't even know the size of $\sccomp(v)$, but it needs to provide this size to the gossiping protocol.
To get around this issue, $v$ runs in parallel $O(\log n)$ copies of a gossiping
protocol for sizes that are powers of $2$. One other challenge is that $v$ needs to be able to determine
whether at least one of these parallel gossiping protocols successfully completed. To achieve this, these gossiping
protocols, in addition to rumors, distribute additional information about the node labels and their in-neighbors. 


\bigskip

\myparagraph{Protocol~{\algSccGossip} for gossiping}.
We will refer to the gossiping algorithm from~\cite{Gasieniec_etal_det_gossip_04} as {\algSccGossip}. 
The following property of {\algSccGossip} is crucial for our algorithm:
\begin{description}
	\item{(scc)} If the input digraph is strongly connected and has at most $k$ vertices, with the node labels from 
		the set $[K] = \braced{0,1,...,K-1}$, then
			algorithm {\algSccGossip} completes gossiping in time $O(k^{4/3}\log K\log^{7/3} k)$. 
\end{description}

As explained earlier, one idea of our algorithm is to execute~{\algSccGossip} on its sc-components.
The details of this will be provided shortly. For now, we only make an observation that captures one 
basic principle of this process. Let $A$ be an sc-component of size $n_A$ and let $j$ be such that $2^{j-1} < n_A \le 2^j$.
Let ${\algSccGossip}_j$ denote {\algSccGossip} specialized for strongly connected
digraphs of size $2^j$ and label set $[n]$,
and let $\sccruntime(j)$ be the running time of ${\algSccGossip}_j$ on such digraphs.
Suppose also that all nodes in $\ingraphinG(A)\setminus A$ are dormant and that the nodes in $A$ execute 
${\algSccGossip}_j$, all starting at the same time.
Since there is no interference from outside $A$, using property (scc) with $k = 2^j$ and $K = n$,
this execution of ${\algSccGossip}_j$ will complete correctly
in the subgraph of $G$ induced by $A$ in time $\sccruntime(j) = \tildeO(n_{A}^{4/3})$. 


\myparagraph{Algorithm~{\algArbitraryGather}}.
Our protocol can be thought of as running two parallel subroutines, the {\SCC}-subroutine  
and the {\ACY}-subroutine, that use two disjoint sets of frequencies. 
There will be $\theta$ \emph{{\ACY}-frequencies} indexed $0,1,...,\theta-1$,
where $\theta = \half(\log n - \log\log n)+2$, as in Section~\ref{sec: tildeO(n1.5) protocol acyclic}. These will
be used by the {\ACY}-subroutine to simulate protocol {\algAcyclicGather}.
We will also have $\theta' = \log n$ \emph{{\SCC}-frequencies} indexed $0,1,...,\theta'-1$, used
by the {\SCC}-subroutine to simulate protocol ${\algSccGossip}$. 
Due to using different frequencies, there will be no signal interference betweeen these two subroutines.


\smallskip
\emparagraph{The {\SCC}-subroutine}.
This subroutine uses the {\SCC}-frequencies, with the {\SCC}-frequency $j$
used to simulate protocol ${\algSccGossip}_j$, for $j=0,1,...,\theta'-1$. For each {\SCC}-frequency $j$, 
any node $v$ divides its time steps into \emph{$j$-frames}, where the $s$-th $j$-frame, for $s = 0,1,...$, 
is $[s\sccruntime(j) , (s+1)\sccruntime(j))$ --- an interval sufficient for a complete
simulation (described below) of ${\algSccGossip}_j$ on a digraph with $2^j$ nodes. 
For each $j$, these simulations start at time $0$ and continue until $v$ determines that for
at least one frequency $j'$ some simulation successfully completed in $\sccomp(v)$.

The overall goal of $v$ executing its {\SCC}-subroutine is to determine $\sccomp(v)$ and
collect all rumors from it. The challenge is that, while $v$ executes its {\SCC}-subroutine,
it may be receiving messages from its in-neighbors \emph{in preceding sc-components}, thus from outside $\sccomp(v)$.
These messages are of two types: ``good'' messages received on
{\ACY}-frequencies, that contain rumors from the in-graph of $v$ and do not interfere with the
{\SCC}-subroutine in $v$, and  ``bad'' messages received on {\SCC}-frequencies that can cause the
{\SCC}-subroutine in $v$ to fail.

We now describe $v$'s simulation of ${\algSccGossip}_j$ on frequency $j$.
The purpose of this simulation is two-fold: one, to determine $\sccomp(v)$, and two, to
distribute all rumors already gathered in $v$ to all nodes in $\sccomp(v)$. 
This is done in two consecutive $j$-frames.  For each $r = 0,1,...$, in the $2r$-th $j$-frame $v$ executes
${\algSccGossip}_j$, using its own label $v$ as the ``rumor'' for the purpose of gossiping.
Let $\tildesccomp(v)$ denote the set of labels received by $v$ during this $j$-frame, including $v$ itself.
In the $(2r+1)$-th $j$-frame, $v$ again executes ${\algSccGossip}_j$, but this
time its ``rumor'' is the vector $[v,\tildesccomp(v),\inneighb(v),\tildeacginneighb(v),R(v)]$, where
$\tildeacginneighb(v)\subseteq \inneighb(v)$ is the set of in-neighbors of $v$ that have
transmitted a message to $v$ on some {\ACY}-frequency (and thus are in a preceding sc-component)
before time  $2r\sccruntime(j)$, and $R(v)$ is the set of all (original) rumors
received on {\ACY}-frequencies before time  $2r\sccruntime(j)$, plus the rumor of $v$. 
(Recall that time step $2r\sccruntime(j)$ is the beginning of $2r$-th $j$-frame.)
Let $\tildesccomp'(v)$ be the set of node labels received in the $(2r+1)$-th $j$-frame.
Then, right after the $(2r+1)$th $j$-frame, $v$ performs three tests:
\begin{description}
	\item{\emph{Test~1:}} Is it true that $\tildesccomp(v) = \tildesccomp(u)$ for all $u\in \tildesccomp(v)$?
	\item{\emph{Test~2:}} Is it true that $\tildesccomp(v) = \tildesccomp'(v)$?
	\item{\emph{Test~3:}} Is it true that  
					$\inneighb(u)\setminus \tildeacginneighb(u) \subseteq \tildesccomp(v)$ for all $u\in \tildesccomp(v)$?  
\end{description}
If one of these tests fails, $v$ continues the execution of the {\SCC}-subroutine. 
If all tests pass, $v$ aborts its {\SCC}-subroutine, discontinues
using all {\SCC}-frequencies, and switches to the {\algAcyclicGather} subroutine, with its
set of collected rumors being $\bigcup_{u\in \tildesccomp(v)} R(u)$.

\bigskip

Unlike in {\algAcyclicGather}, with each node $v$ we now associate \emph{two} activation times.
The first one is called $v$'s \emph{SCC-activation} and
is defined analogously to the activation time in {\algAcyclicGather}: If $\acyinneighb(v) = \emptyset$ then $\sccacttime(v)= 0$.
Otherwise,  $\sccacttime(v)$ is the last-received value $\rws^1(u,v)$ for $u\in \acyinneighb(v)$, 
where  $\rws^1_{u,v}$ denotes the first  $\rws_u$ value received by $v$ from $u$.   
As explained earlier, these values will be received on the {\ACY}-frequency.  
(Note that
the algorithm does not actually use SCC-activation values for computation --- these will be used only for the analysis.)
If $r$ is the index such that Tests~1 and~2 pass after the double $j$-frames
$2r$ and $2r+1$ then the second activation time for $v$ is $\acyacttime(A) = (2r+2)\sccruntime(j)$.


\smallskip
\emparagraph{The {\ACY}-subroutine}.
We refer to the value $\acyacttime(v)$ defined above as $v$'s \emph{ACG-activation time}. 
This value now plays the role of $v$'s activation time in protocol {\algAcyclicGather}.
In this subroutine $v$ will transmit at the {\ACY}-frequencies and $v$ simply
executes {\algAcyclicGather}, starting at time $\acyacttime(v)$, in its activity period 
$[\acyacttime(v),\acyacttime(v)+\beta_\theta)$. The activity stages and the transmissions of each 
node are defined in exactly the same way as in protocol {\algAcyclicGather}
(except that we use  $\acyacttime(v)$ instead of $\acttime(v)$). 


\myparagraph{Correctness}.
We justify correctness first. Note that any node $v$ is guaranteed to successfully transmit 
during the {\ACY}-subroutine, because this subroutine involves a round of {\RoundRobin}.
Thus it suffices to prove that each node $v$ correctly completes the {\SCC}-subroutine,
meaning that it will eventually correctly compute $\tildesccomp(v) = \sccomp(v)$ and stop the {\SCC}-subroutine.

The proof of this property is by induction on the size of $v$'s in-graph $\ingraphinG(v)$. Assuming that all
nodes in $\ingraphinG(v)$ satisfy this property, we argue that it also holds for $v$.
First, we show that if $v$ stops its {\SCC}-subroutine then $\tildesccomp(v) = \sccomp(v)$. Indeed,
Tests~1-2 imply that each $u\in \tildesccomp(v)$ and $v$ are reachable from each other, and therefore $\tildesccomp(v)\subseteq\sccomp(v)$.
And if we had $\sccomp(v)\setminus \tildesccomp(v)\neq\emptyset$ then there would be a vertex in $\sccomp(v)\setminus \tildesccomp(v)$
with an out-neighbor $u$ in $\tildesccomp(v)$, contradicting Test~3. So, as long as the {\SCC}-subroutine of $v$
completes, we have $\tildesccomp(v) = \sccomp(v)$. On the other hand,
the paragraph before the description of the algorithm shows that after all nodes in $\ingraphinG(v)\setminus\sccomp(v)$
complete their {\SCC}-subroutines correctly, and thus cease using {\SCC}-frequencies, if $v$ still has not
completed its {\SCC}-subroutine, then it will correctly compute $\tildesccomp(v) = \sccomp(v)$ and it
will have all rumors from $\ingraphinG(v)$.
 

\myparagraph{Running time}.
Next, we estimate the running time. The argument follows the reasoning in Section~\ref{sec: tildeO(n1.5) protocol acyclic},
but now we need to account for the contribution of the {\SCC}-subroutine. 
The idea was already explained in the paragraph before the description of the algorithm, where  we show that the delay caused by
the need to distribute rumors in an sc-component $A = \sccomp(v)$ of $v$ is only $\tildeO(n_A^{4/3})$, and thus less than $\tildeO(n_A^{1.5})$, and so we can charge this delay to the nodes in $A$. A more formal argument follows.

When a node $v$ starts its {\ACY}-subroutine at time $\acyacttime(v)$, the {\SCC}-subroutine
in  $A = \sccomp(v)$ has already completed. By applying this property to 
the nodes in $\acyinneighb(v)$, we obtain that when $v$ starts its {\SCC}-subroutine at time $\sccacttime(v)$, the {\SCC}-subroutines in
all sc-components in the in-graph $\ingraphinG(v)$ of $v$ have already completed.  
Let $\sccacttime(A) = \max_{u\in A} \sccacttime(u)$. By the earlier
observation, all nodes in $A$ will already have all rumors from the in-graph $\ingraphinG(A)$
at time $\sccacttime(A)$, and therefore  the execution of {\algSccGossip} in $A$ will be successful in the
SCC-frame starting at $\sccacttime(A)$. This implies that   $\acyacttime(A) = \sccacttime(A) + 2\cdot\sccruntime(A)$. 

The above paragraph implies that, for the nodes in $A$, the contribution per node of
the {\SCC}-subroutine to the overall running time is at most $2\sccruntime(A)/n_A = \tildeO(n^{1/3})$.
The analysis of the {\ACY}-subroutine is the same as for protocol  {\algAcyclicGather}, giving us that
its contribution per node to the overall running time is $\tildeO(n^{1/2})$. These two
facts imply the $\tildeO(n^{1.5})$ upper bound on the running time of Algorithm~{\algArbitraryGather}.

To make this argument more precise, we extend the definition of a critical path from 
Section~\ref{sec: tildeO(n1.5) protocol acyclic}.
In this section, the critical path is a sequence of nodes $v_0w_0v_1w_1 ... v_pw_p=t$  defined as follows:
\begin{itemize}
	   \item For each $a= p,p-1,...,0$, suppose that $w_a$ has already been defined, and let $C_a = \sccomp(w_a)$.
			If $\bigcup_{u\in C_a} \acyinneighb(u)\neq\emptyset$, then let
				$v_a \in A$ be the node for which  $\sccacttime(v_a) = \sccacttime(C_a)$.
				In other words, $v_a$ is the node in $C_a$ for which   $\sccacttime(v_a)$ is maximum. 
				(It could happen that $v_a = w_a$.)  On the other hand, if $\bigcup_{u\in C_a} \acyinneighb(u) = \emptyset$
			  (that is, $C_a$ is a source sc-component), then $a=0$ and
			   $v_0\in C_a$ is arbitrary, for example we can take $v_0  = w_0$.
		\item  For each $a = p-1,p-2,...,0$, suppose that $v_{a+1}$ has
				already been defined.  Then $w_a$ is the node in $\acyinneighb(v_{a+1})$
				whose message was received last by $v_{a+1}$ (formally,  $w_a$ is chosen so that $\sccacttime(v_{a+1}) = \rws^1(w_a,v_{a+1})$).
\end{itemize}


\begin{figure}[ht] 
\begin{center}
\includegraphics[width = 4.5in]{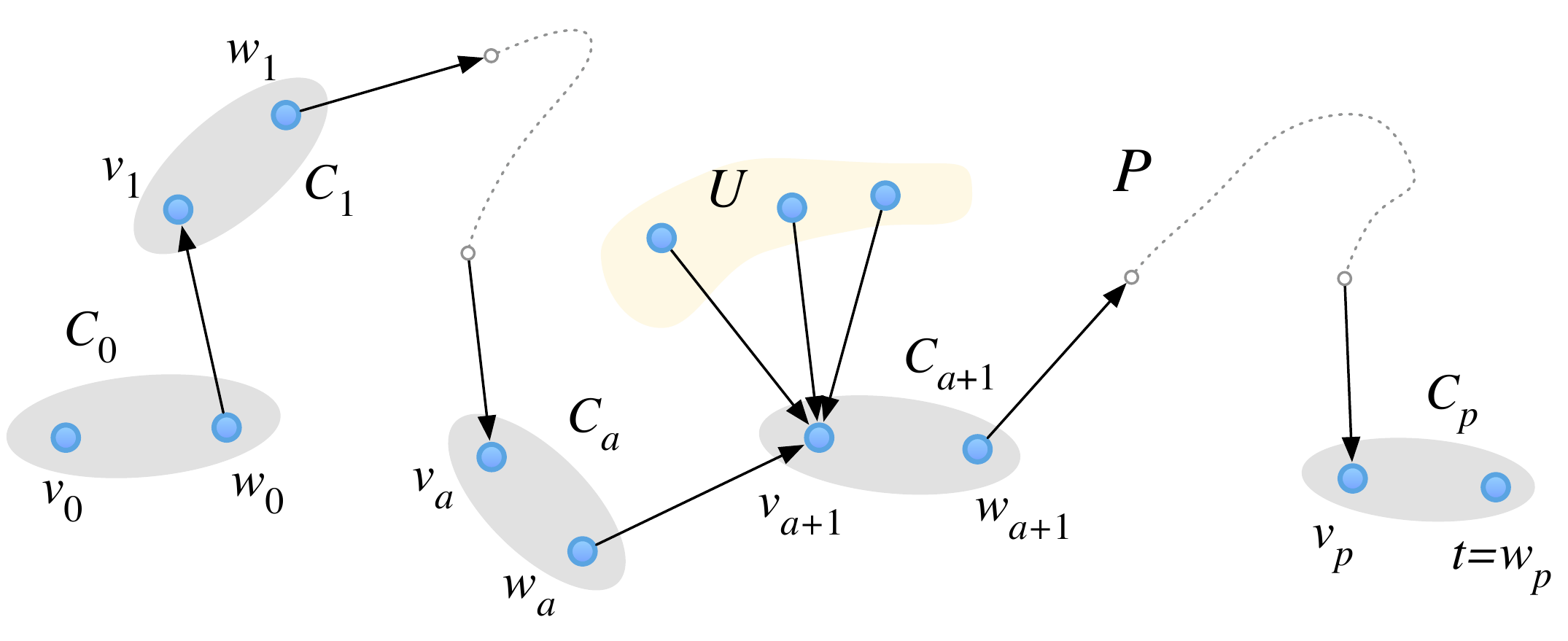} 
\end{center}   
\caption{Illustration of the time analysis for arbitrary graphs.} 
\label{fig: time analysis arbitrary}     
\end{figure}


Denote by $T$ the running time of protocol~{\algArbitraryGather}. We have $T \le \acyacttime(C_p)$ and $\sccacttime(C_0) = 0$, 
so we can the express $T$ as
\begin{align*}
	T\; &=\; \sum_{a = 0}^p  \,[\, \acyacttime(C_a) - \sccacttime(C_a) \,] 
				+ \sum_{a = 0}^{p-1} \, [\, \sccacttime(C_{a+1}) - \acyacttime(C_a) \,]  \,.
\end{align*}
We estimate the two terms separately. As explained earlier, we have $\acyacttime(C_a) = \sccacttime(C_a)+ 2\cdot\sccruntime(C_a)$, 
so the first term is at most 
\begin{align*}
\sum_{a = 0}^p \, [\, \acyacttime(C_a) - \sccacttime(C_a) \,] \;&=\; 2\cdot \sum_{a = 0}^p \sccruntime(C_a)
\\
		&=\;  \sum_{a = 0}^p \tildeO(n_{C_a}^{4/3})	
		\;=\; \tildeO(n^{4/3}),
\end{align*}
because $\sum_{a=0}^p n_{C_a} \le n$. To estimate the second term, note that 
the definition of $v_{a+1}$ implies that $\sccacttime(C_{a+1}) = \sccacttime(v_{a+1})$.
Further, in the execution of {\algAcyclicGather}, node $w_a$ gets activated at time $\acyacttime(C_a)$.
Then the analysis identical to that in Section~\ref{sec: tildeO(n1.5) protocol acyclic}
yields that we can estimate the second term by $\tildeO(n^{1.5})$.

As in the previous section, a more accurage bound follows by observing that
in the analysis above we ignored three $\log n$ factors. We thus obtain the main result of this paper:


\begin{theorem}\label{lem: running time of  ArbGather}
Let $G$ be an arbitrary digraph with $n$ vertices and a designated target node reachable from all other nodes.
Algorithm~{\algArbitraryGather} completes information gathering in $G$ in time $O(n^{1.5}\log^3 n)$.	
\end{theorem}


\section{$\tildeO(n)$-Time Protocol With Acknowledgements for Acyclic Graphs} 
\label{sec: protocol with acknowledgement}


We now consider the problem of gathering in acyclic graphs with a weak form of acknowledgment of transmission success.  
To be more precise: Following each transmission from a node $v$, $v$ receives a single bit indicating whether at least 
one node successfully received that transmission ($v$ does not learn which specific node, or how many nodes in total, 
received the transmission).  Our main goal in this section is to show that this single bit is enough to allow for 
gathering to be performed in time $O(n \log^2 n)$ on acyclic graphs with $n$ vertices.The key idea here will be that 
nodes which have successfully transmitted can at least temporarily stop transmitting, making it easier for other nodes 
to succeed.  In order for this to work, though, we need to guarantee that successful transmissions are occurring at 
a reasonable rate.  The following combinatorial object will be our main tool for this.

We say that a collection $(S_0, S_1, \dots, S_{\ell-1})$ of label sets forms a \emph{$(n,k)$-half-selector} if for every $X \subseteq [n]$ 
with $|X| \le k$ there are at least $|X|/2$ choices of  $x \in X$ for which there is an index $i$ with $S_i \cap X = \{x\}$ 
(in contrast to strong selectors where we want this property to hold for every choice of $x$).  
It is a consequence of Lemma~1 in~\cite{Chrobak_etal_fast_02} that for every $k$ there exists a half-selector of size $O(k \log n)$.

For all $j=0,1,\dots,\log n$, by $\HalfSelector{2^j}=(S_0^j, S_1^j, \dots, S_{b_j-1})$ we will denote a $2^j$-half-selector
of size $b_j=O(2^j \log n)$.  Without loss of generality we can also assume that $b_{j+1}=2b_j$ for all $i \leq \log n-1$, 
implying that $b_j=\gamma 2^j \log n$ for some absolute constant $\gamma$.

As in the previous sections, our algorithm will run on multiple frequencies, though this time the number of frequencies 
is $\kappa=\log n+2$.  The intuition here is that for $0 \leq j \leq \kappa-2$ frequency $j$ will be used to handle 
potential interferences involving at most $2^j$ vertices.


\myparagraph{Protocol~{\algAcyclicGatherWithAck}.}
At any given time step, a node can be either \textit{dormant} or \textit{active}. Initially the source nodes (with no in-neighbors) 
will be active and the remaining nodes will be dormant. Any active node
transmits according to $\HalfSelector{2^j}$ on each frequency $j = 0, 1, ..., \kappa-2$, 
and according to {\RoundRobin} on frequency $\kappa-1$.  
An active node which receives an acknowledgement of a successful transmission moves to the dormant state, and a dormant 
node which receives a transmission becomes active.  Note that, unlike the 
previous algorithms, it is now possible for a node to become active multiple times during the process as it continually 
receives new rumors.

\myparagraph{Correctness.}
As in the previous algorithms, correctness follows immediately from the inclusion of \RoundRobin.

\myparagraph{Running time.}
We claim that the running time of this protocol (with $\kappa$ frequencies) is $O(n\log n)$. 
Since $\kappa = O(\log n)$, this will give us an $O(n\log^2n)$-time protocol in the standard single-frequency model.

For a given node $v$, let $\maxpath(v)$ denote the length of the \textit{longest} directed path from $v$ to the target node 
$t$. (This path cannot repeat vertices due to our assumption that $G$ is acyclic.)  
Let $\maxmaxpath = \max_{v\in G} \maxpath(v)$. For $i = 0,1,...,\maxmaxpath$, let $B_i$ denote the set of nodes with 
$\maxpath(v)=\maxmaxpath-i$.
(So $B_\maxmaxpath = \braced{t}$ and $B_0$ consists of the nodes with the longest path to $t$).  
The following observation is immediate from the definition of $B_i$'s:

\begin{observation}\label{obs:layers}
If $i < i'$ then there are no edges from $B_{i'}$ to $B_i$.
In particular, the vertices in $B_0$ have no incoming edges.
\end{observation}

Let $\tau_{i}= 4 \gamma \sum_{p < i}|B_i| \log n$ for all $i$. (In particular, $\tau_0 = 0$.)
Our running time bound would follow from the following claim: 

\begin{claim}\label{obs:ackcomplete}
The following two properties hold for every $i = 0,1, ...,\maxmaxpath$:  
\begin{description}
\item{(i)} All nodes in $\bigcup_{p < i} B_p$ remain dormant at all times after $\tau_i$ (inclusive).
\item{(ii)} At time $\tau_i$ each rumor is in $\bigcup_{p \ge i} B_p$. 
\end{description}
In particular, at time $\tau_{\maxmaxpath} = 4\gamma \sum_{p=0}^{r-1} |B_p| \log n < 4 \gamma n \log n$ 
each rumor will be $t$.
\end{claim}

We establish Claim~\ref{obs:ackcomplete} inductively.  Both parts~(i) and~(ii)
of the claim hold vacuously for $i=0$.  
Now suppose the claim is true for some $i$ and consider the computation of the nodes in $B_i$ beginning at time $\tau_{i}$.  
These nodes will not receive any rumors after time $\tau_{i}$ since, by the inductive hypothesis~(i) and Observation~\ref{obs:layers},
none of their in-neighbors will be active at any point. So any node in $B_i$ already dormant at time $\tau_{i}$ remains 
dormant, and any active node in it becomes permanently dormant once it succeeds at least once.

Choose $j$ such that $2^{j-1} < |B_i| \le 2^{j}$. Let $A$ be the set of nodes in $B_i$ that are active
at time $\tau_i$. Trivially, $|A|\le |B_i| \le 2^j$.
Since the algorithm runs $\HalfSelector{2^j}$ on frequency $j$,
at least $|A|/2$ nodes in $A$ will have a time step in the interval $[\tau_i, \tau_i+b_j)$
when they will successfully transmit and become dormant. Thus, if $A'$ is the set of nodes in $B_i$ that
are active at time $\tau_i + b_j$, then $|A'|\le |A|/2 \le 2^{j-1}$.
Next, we look at time interval $[\tau_i+b_j, \tau_i+b_j+b_{j-1})$. 
Since the algorithm runs $\HalfSelector{2^{j-1}}$ on frequency $j-1$, using the same argument,
if $A''$ is the set of nodes in $B_i$ active at time  $\tau_i+b_j+b_{j-1}$ then $|A''|\le |A|/2 \le 2^{j-2}$. 
Continuing inductively, 
all the nodes in $A$ will succeed and become dormant no later than at time
\begin{align*}
	\textstyle
	\tau_i+ \sum_{q=0}^{j} b_q \;&=\; \textstyle \tau_i+\gamma (\sum_{q=0}^{j} 2^q) \log n 
				\\
	\textstyle						&<\; \tau_i + \gamma 2^{j+1} \log n 
				 \\
							&\leq\; \tau_i + 4 \gamma |B_i| \log n 
	  \textstyle   					\;=\; \tau_{i+1}.
\end{align*}
Thus all the nodes in $B_i$ become dormant by time $\tau_{i+1}$ and will stay dormant, showing~(ii). 
By Observation~\ref{obs:layers}, each successful transmission from $B_i$ arrives at a node in $\sum_{p\ge i+1} B_p$, 
so part~(ii) is also established. Concluding, we have proved the following theorem.


\begin{theorem}\label{lem: running time of AcyclicGather with ack}
Let $G$ be an acyclic directed graph with $n$ vertices and a designated target node reachable from
all other nodes. Using acknowledgements of successful transmissions,   
Algorithm~{\algAcyclicGatherWithAck} completes information gathering in $G$ in time $O(n\log^2 n)$.
\end{theorem}


\section{Final Comments}
\label{sec: final comments}



In this paper we provided an $\tildeO(n^{1.5})$-time protocol for 
information gathering in ad-hoc radio networks, improving the trivial upper bound of $O(n^2)$.
For the model with transmissions acknowledgments
we gave a $\tildeO(n)$-time protocol for acyclic graphs. 

We hope that some ideas behind our algorithms will lead to further improvements, and
perhaps find applications to other communication dissemination problems in ad-hoc radio networks. 
One idea that is particularly promising is the 
amortization technique in Section~\ref{sec: tildeO(n1.5) protocol acyclic},
where a failure of a node in transmitting its message is charged to stage indices
of the interfering nodes. Another idea is the technique for integrating a gossiping
protocol (applicable only to strongly connected digraphs) with an information gathering protocol
for acyclic digraphs, to obtain an information gathering protocol for arbitrary digraphs.
Using this technique, improving the upper bound to below $\tildeO(n^{1.5})$ should be
possible by designing an appropriate protocol for the restricted case of acyclic graphs.

Several open problems remain. The two most intriguing problems are about the
time complexity of gossiping and information gathering, as for both problems
the best known lower bounds are only $\Omega(n\log n)$, the same as for broadcasting.


\bibliographystyle{plain}
\bibliography{information_gathering}

\end{document}